\documentstyle [12pt] {article}
\title{MPEMBA EFFECT, NEWTON COOLING LAW AND HEAT TRANSFER EQUATION}

\author{Vladan Pankovi\'c, Darko V. Kapor\\
Department of Physics, Faculty of Sciences, 21000 Novi Sad,\\ Trg
Dositeja Obradovi\'ca 4, Serbia, \\vladan.pankovic@df.uns.ac.rs}

\date {}
\begin {document}
\maketitle \vspace {0.5cm}
\vspace {1.3cm}

\begin {abstract}
In this work we suggest a simple theoretical solution of the
Mpemba effect in full agreement with known experimental data. This
solution follows simply as an especial approximation
(linearization) of the usual heat (transfer) equation, precisely
linearization of the second derivation of the space part of the
temperature function (as it is well-known Newton cooling law can
be considered as the effective approximation of the heat
(transfer) equation for constant space part of the temperature
function).
\end {abstract}

\vspace {1.0cm}

In this work we shall suggest a simple theoretical solution of the
Mpemba effect in full agreement with known experimental data
[1]-[4]. This solution follows simply as an especial approximation
(linearization) of the usual heat (transfer) equation, precisely
linearization of the second derivation of the space part of the
temperature function (as it is well-known Newton cooling law [5]
can be considered as the effective approximation of the heat
(transfer) equation for constant space part of the temperature
function).

As it is well-known [5] usual Newton's law of cooling represents
the following simple linear differential equation
\begin {equation}
    \frac {dT}{dt} = - {\alpha}(T-T_{e})
\end {equation}
with simple solution
\begin {equation}
     T = (T_{0}- T_{e})\exp[-{\alpha}t]  + T_{e}               .
\end {equation}
Here $T$ represents the temperature of a liquid with initial value
$T_{0}$, $T_{e}$ - constant, smaller than $T$, temperature of a
cooling wall (thermostat or environment) in the contact with
liquid, while ${\alpha}$ represents a cooling parameter that
simplifiedly expresses thermodynamics of the cooling process. It
is supposed implicitly that cooling parameter is independent of
the initial temperature of the liquid as well as of the cooling
wall temperature.

Consider now described by (1) or (2) thermodynamical interaction
between liquid and cooling wall in two different cases, precisely
for two different initial temperatures of the liquid, first one
$T_{10}$ and second one $T_{20}$ so that $T_{10}$  is larger than
$T_{20}$  , i.e.
\begin {equation}
      T_{10}> T_{20}                                         .
\end {equation}
Then (2), (3) imply
\begin {equation}
     \frac {T_{1}-T_{e}}{ T_{2}-T_{e} }=\frac {T_{10}-T_{e}}{ T_{20}-T_{e} }
\end {equation}
where $T_{1}$ and $T_{2}$ represents the temperature of the liquid
in the first case and second case in some (finite) later time
moment t. It means that temperature of the liquid in the first
case will be larger than the temperature of the liquid in the
second case in any (finite) later time moment.

Previous conclusion implies unambiguously that usual Newton's
cooling law (1) cannot model experimental data of the Mpemba
effect [1]-[4] according to which, simply speaking, hot liquid
freezes faster than cold liquid.

Consider, however, the standard (one-dimensional) heat transfer
differential equation
\begin {equation}
      \frac {\partial^{2}T}{\partial x^{2}}= k \frac {\partial T}{\partial t}                   .
\end {equation}
Here $x$ represents the coordinate (along $x$-axis) between 0 and
linear dimension of the vessel $L>0$, $t$ - time moment between 0
and $\infty$, and $k$ - corresponding (positive) constant, so that
liquid temperature $T$ represents the function of $x$ and $t$.
This equation, of course, represents mathematically partial
differential equation of the second range.

Suppose additionally that $T$ satisfies the following two, initial
and "final" conditions,
\begin {equation}
  T(x,0) = T_{0}
\end {equation}
and
\begin {equation}
   T(x,t) = T_{e}
\end {equation}
for $t\gg 0$ .

As it is well-known solution of heat equation (5) can be supposed
in the following form
\begin {equation}
   T(x,t) = X(x)\tau(t) + A
\end {equation}
where $X(x)$ represents the space part of $T(x,t)$ depending of
$x$ only, $\tau(t)$ - time part of $T(x,t)$ depending of $t$ only
and $A$ some constant.

Introduction of (8) in (5) yields
\begin {equation}
      \tau \frac {d^{2}X}{dx^{2}}= k X \frac {d\tau}{dt}
\end {equation}
and further
\begin {equation}
   \frac {1}{X}\frac {d^{2}X}{dx^{2}} = \frac {k}{\tau}\frac {d\tau}{dt} = - \beta
\end {equation}
where $\beta$ represents some (positive) constant. It, in fact,
represents a system of two usual differential equations with
simple solutions
\begin {equation}
    \tau = B\exp [-\frac {\beta}{k}t]
\end {equation}
\begin {equation}
     X = C \cos [\beta^{\frac {1}{2}} x] + D \sin [\beta^{\frac {1}{2}}x]
\end {equation}
where $B$, $C$ and $D$ are some constants.

Then, according to (8), it follows
\begin {equation}
   T = \exp [-\frac {\beta}{k}t] (BC \cos [\beta^{\frac {1}{2}} x] + BD \sin [\beta^{\frac {1}{2}}x]) + A
\end {equation}
which, according to initial and "final" conditions (6), (7) that
imply
\begin {equation}
   A = T_{e}
\end {equation}
\begin {equation}
   BC = BD = T_{0} - T_{e}
\end {equation}
, turns out in
\begin {equation}
   T = (T_{0} - T_{e}) \exp [-\frac {\beta}{k}t] (\cos [\beta^{\frac {1}{2}}x] + sin [\sin [\beta^{\frac {1}{2}}x]) + T_{e}           .
\end {equation}
Obviously, in an approximation in which $(\cos [\beta^{\frac
{1}{2}}x] + sin [\sin [\beta^{\frac {1}{2}}x])$  can be
approximately treated as a constant, i.e. average value of this
expression close to 1, solution (16) can be effectively
approximated by the solution of the Newton cooling law (2). Simply
speaking, Newton cooling law can be considered as the effective
approximation of the heat (transfer) equation for constant space
part of the temperature function. But it implies that standard
heat (transfer) equation, as well as Newton cooling law, cannot
model experimental data of the Mpemba effect.

Consider, now, such situation in which temperature function
changes very slowly by space coordinate change so that the
following linearization of the temperature second space derivation
can be approximately realized
\begin {equation}
      \frac {\partial^{2} T}{\partial x^{2}} \simeq \frac {T_{0} - T_{e}}{L} \frac {\partial T}{\partial x}                 .
\end {equation}
Such and similar kind of approximation is very often in different
important domains of the physics, e.g. quasi-classical
approximation of the quantum mechanics, i.e. WBK formalism, solid
state physics, quantum field theory, inflationary cosmology etc. .

In such approximation standard heat (transfer) equation turns out
in
\begin {equation}
    \frac {T_{0} - T_{e}}{L} \frac {\partial T}{\partial x}= k \frac {\partial T}{\partial t}
\end {equation}
or, after application of (8), in
\begin {equation}
     \frac {1}{X}\frac {dX}{dx}= \frac { kL }{T_{0} - T_{e}} \frac {1}{\tau}\frac {d\tau}{dt} = - \beta    .
\end {equation}
It, in fact, represents a system of two usual differential
equations with simple solutions
\begin {equation}
    \tau = B \exp[- \beta \frac { T_{0} - T_{e}}{kL} t]
\end {equation}
\begin {equation}
  X = C \exp [-\beta x]
\end {equation}
where $B$, $C$ are some constants.

Then, according to (8), it follows
\begin {equation}
    T = BC \exp [-\beta x] \exp[- \beta \frac { T_{0} - T_{e}}{kL} t] + A
\end {equation}
which, according to initial and "final" conditions (6), (7) that
imply
\begin {equation}
   A = T_{e}
\end {equation}
\begin {equation}
  BC = T_{0} - T_{e}
\end {equation}
, turns out in
\begin {equation}
  T = (T_{0} - T_{e}) \exp [-\beta x] \exp[- \beta \frac { T_{0} - T_{e}}{kL} t]  + T_{e}         .
\end {equation}
Obviously here $T$ decreases exponentially during time with
exponent parameter proportional to $ T_{0} - T_{e}$. It means
that, simply speaking, hot liquid can freeze faster than cold
liquid in full agreement with experimental data of the Mpemba
effect.

Precisely, consider two equivalent liquids in the same vessels so
that
\begin {equation}
     T_{20} > T_{10}
\end {equation}
where $ T_{20}$  represents the larger initial temperature of the
second liquid while $ T_{10}$ represents the smaller initial
temperature of the first liquid. Suppose that the first liquid
obtains some temperature $T$ (25) in a time moment $t_{1}$ while
the second liquid obtains the same temperature $T$ in some other
time moment $t_{2}$. Then simple calculations, according
suppositions and (25), yield
\begin {equation}
    t_{1} > \frac { T_{20} - T_{e}}{ T_{10} - T_{e}} t_{2} > t_{2}
\end {equation}
corresponding to Mpemba effect experimental data.

In this way it is shown that the suggested approximation of the
heat transfer equation can successfully described the Mpemba
effect experimental data. Moreover it follows that the Mpemba
effect can appear only in situations when given approximation is
satisfied, while in opposite situations Mpemba effect does not
appear. Finally, according to (25), in situations when Mpemba
effect appears this effect becomes sharply expressed when $L$ is
smaller in full agreement with experimental data on the Mpemba
effect [1]-[4].

In conclusion it can be repeated and pointed out the following. In
this work we suggest a simple theoretical solution of the Mpemba
effect in full agreement with known experimental data. This
solution follows simply as an especial approximation
(linearization) of the usual heat (transfer) equation, precisely
linearization of the second derivation of the space part of the
temperature function (as it is well-known Newton cooling law can
be considered as the effective approximation of the heat
(transfer) equation for constant space part of the temperature
function).

This work is dedictated to memory on Prof. Dr. Bratislav Tosic. He
stated, we paraphrase, approximation is not only ground tool but
holly spirit too of any physical theory.

\vspace {1.3cm}

{\large \bf References}

\vspace {1.3cm}

\begin {itemize}

\item [[1]] E. B. Mpemba, D. G. Osborne, Phys. Educ. {\bf 4} (1969) 172
\item [[2]] D. Auerbach, A. J. Phys. {\bf 63} (1995) 882
\item [[3]] M. Jeng, A. J. Phys. {\bf 74} (2006) 514
\item [[4]] J. I. Katz, Am. J. Phys. {\bf 77} (2009) 27
\item [[5]] C. T. O'Sullivan, Am. J. Phys. {\bf 58} (1990) 956

\end {itemize}

\end {document}